**Climbing, falling and jamming during ant locomotion in confined environments**

Nick Gravish[1], Daria Monaenkova[1], Michael A. D. Goodisman[2], and Daniel I. Goldman[1]

[1]School of Physics, Georgia Institute of Technology, [2]School of Biology, Georgia Institute of Technology

Corresponding author: daniel.goldman@physics.gatech.edu

**Abstract**

Locomotion emerges from effective interactions of an individual with its environment. Principles of biological terrestrial locomotion have been discovered on unconfined vertical and horizontal substrates. However a diversity of organisms construct, inhabit, and move within confined spaces. Such animals are faced with locomotor challenges including limited limb range of motion, crowding, and visual sensory deprivation. Little is known about how these organisms accomplish their locomotor tasks, and such environments challenge human-made devices. To gain greater insight into how animals move within confined spaces we study the confined locomotion of the fire ant *Solenopsis invicta*, which constructs subterranean tunnel networks (nests). Laboratory experiments reveal that ants construct tunnels with diameter, D, comparable to bodylength, L=3.5 ± 0.5 mm. Ants can move rapidly (> 9 bodylengths/sec) within these environments; their tunnels allow for effective limb, body, and antennae interaction with walls which facilitate rapid slip-recovery during ascending and descending climbs. To examine the limits of slip-recovery in artificial tunnels we perform perturbations consisting of rapid downward accelerations of the tunnels, which induce falls. Below a critical tunnel diameter, $D_s$=1.31 ± 0.02 L, falls are always arrested through rapid interaction of appendages and antennae with tunnel walls to jam the falls. $D_s$ is comparable to the size of incipient nest tunnels (D = 1.06 ± 0.23 L) supporting our hypothesis that fire ants construct environments which simplify their control task when moving through the nest, likely without need for rapid nervous system intervention.

*Introduction*

Terrestrial animals and increasingly robots must move in diverse and complex environments, including running across flat landscapes (1), swimming in sand (2), climbing rough or smooth vertical surfaces(3), and squirming through cracks (4). The bulk of discoveries of locomotor behaviors and control strategies have been made by challenging animals in the laboratory in simplified  environments that are typically featureless, flat and unconfined (5). Such simplifications have allowed discovery of general principles in locomotor modes of walking, running and climbing (6-9). Recent studies have generated appreciation for the importance of mechanical interactions with the environment, and through biological experiment (10) and robot modeling (11, 12) have demonstrated that stable and robust movement can emerge as a result of appropriately tuned dynamics of limb-ground interaction (13, 14). For example rapid perturbations to locomotion may be corrected by so-called preflexes (15) in which mechanical design of the limb and appropriate kinematics enable rapid recovery from perturbation (6, 8, 10). However, typical substrates that legged locomotors contend with differ in orientation, can deform in response to foot/body contact (1, 11) and are rough on multiple size scales (16, 17); little is known about how organisms effectively utilize limb/body-substrate interactions in such environments. Practically, we expect that discovery of such principles can lead to advances of robotic devices that must operate in complex conditions; such devices often suffer performance loss in natural environments (18).

In particular, the role of confinement is relatively unexplored in locomotor performance and behavior. Many subterranean-dwelling organisms live and move within *confined* spaces in their environments (19, 20). The morphology(21, 22), energetic costs (23-26), and genetic basis (27, 28) for creating subterranean burrows and nests, which are examples of the "extended phenotype"(29), have been studied across a diversity of organisms. However, the constraints on locomotion of individuals and groups inhabiting these environments are largely unexplored (30). Rapid locomotion within the confines of a subterranean nest is essential for inhabitants to escape or respond to predators (19, 31), evacuate during flooding(32), or transport resources and information effectively (31). However lack of vision (19, 20, 33), limited limb mobility (19), and excessive crowding among individuals (34) would seemingly challenge efforts at rapid locomotion within confined environments. Thus we seek to understand how such environments influence the mobility and stability of animals moving within them.

Ants are excellent organisms with which to study confined locomotion. Many ant species construct large underground nests through the excavation of soil (35). Nest shape and size—in addition to ant shape and size—varies widely across species but typically consists of vertical tunnels that connect larger chambers used for food storage and brood rearing (22, 35). A majority of an ant colony worker's life is spent below the surface within the nest—tending to brood or performing routine nest maintenance—and only near the end of life do worker ants forage above surface (35-38). The evolutionary pressure of subterranean life has led to several adaptations among ants such as partial or complete loss of vision in some species (39, 40) and long-range acoustic (41-43) and chemical communication systems (39, 44, 45). However, almost nothing is known about how ants move through their confined nest environments.

We hypothesize that ants have developed strategies and adaptations for rapid movement within nests, particularly during crucial times such as nest reconstruction or evacuation. A species that frequently must contend with such events is the red imported fire ant (*Solenopsis invicta*). Fire ants originate from the Pantanal wetlands in South America, which are subject to seasonal rains and flooding (38). Fire ant colonies construct large and relatively complex subterranean nests (38) which can be up to 2 m deep and contain greater than 50 m in length of tunnels (46). As an invasive species in the Southern United States fire ants have demonstrated their proficiency at constructing nests within a wide range of soil conditions (38). Construction of such large nests demands the ability to move repeatedly and stably within the nest confines while transporting soil.

In this article we seek to identify principles of locomotion within confined environments which challenge animals with a different set of locomotive constraints than in above-ground study. We investigate the effects of subterranean confinement (tunnel diameter) on the mobility and stability of the fire ant (*Solenopsis invicta*). We show that climbing in confined environments is a robust mode of high-speed locomotion, in which slips, falls, and frequent collisions with the environment do not necessarily prevent high-speed ascent and descent. We also demonstrate an unusual stabilizing response of fire ants when dislodged from the tunnel wall—the use of antennae as limb-like appendages to arrest and jam falls. Overall, we find that stable locomotion within subterranean environments is a function of the local tunnel morphology within which the organisms move. We hypothesize that the principle of off-loading locomotor control to the environment can be used by animals in confined enviroments and can inspire the next generation of mobile robots.

## Results and discussion
*The shape and form of excavated fire-ant tunnels*

To examine the interaction of fire ants with the tunnels that they constructed, we first measured the size and shape of nest tunnels excavated by fire ant workers (body length L = 0.35 ± 0.05, N = 2,611 measurements) in three-dimensions in a laboratory experiment using an X-ray computed tomography (CT) system (Fig. 1). We allowed isolated groups of fire ant workers to excavate tunnels within an 8 cm diameter, and 12 cm deep, cylindrical volume of laboratory soil (wet approximately spherical glass particles, see below) over the course of 20 hours. The tunnels were roughly circular in cross section (Fig. 1b and SI) and the effective cross sectional diameter (See SI) within the tunnels was, D = 3.7 ± 0.8 mm (N = 2,262 observations from 10 experiments).

To determine if the soil-substrate had an effect on tunnel shape and size, we repeated this experiment using different substrate combinations of particle diameter (50, 210, 595 μm; See Table S1 for polydispersity) and soil moisture content (1, 3, 5, 10, 15, 18, 20% by mass). We challenged worker groups from 8 separate colonies to excavate tunnels in each substrate combination and collected 168 separate X-Ray CT tunnel excavation observations (Fig. S3). We found a significant effect of both particle diameter ($F_{2,136}$ = 10.48, p < 0.0001) and soil moisture content ($F_{6,136}$ = 5.38, p < 0.0001) on excavated tunnel depth indicating that substrate had a strong effect on digging proficiency. Soil-moisture content had a non-linear effect on tunnel depth.     Tunnel     depth     was     small     at     low     soil-moisture and rose to a maximum at intermediate soil-moisture contents of 10-15%, above which tunnel depth decreased again at high soil-moisture (See SI).

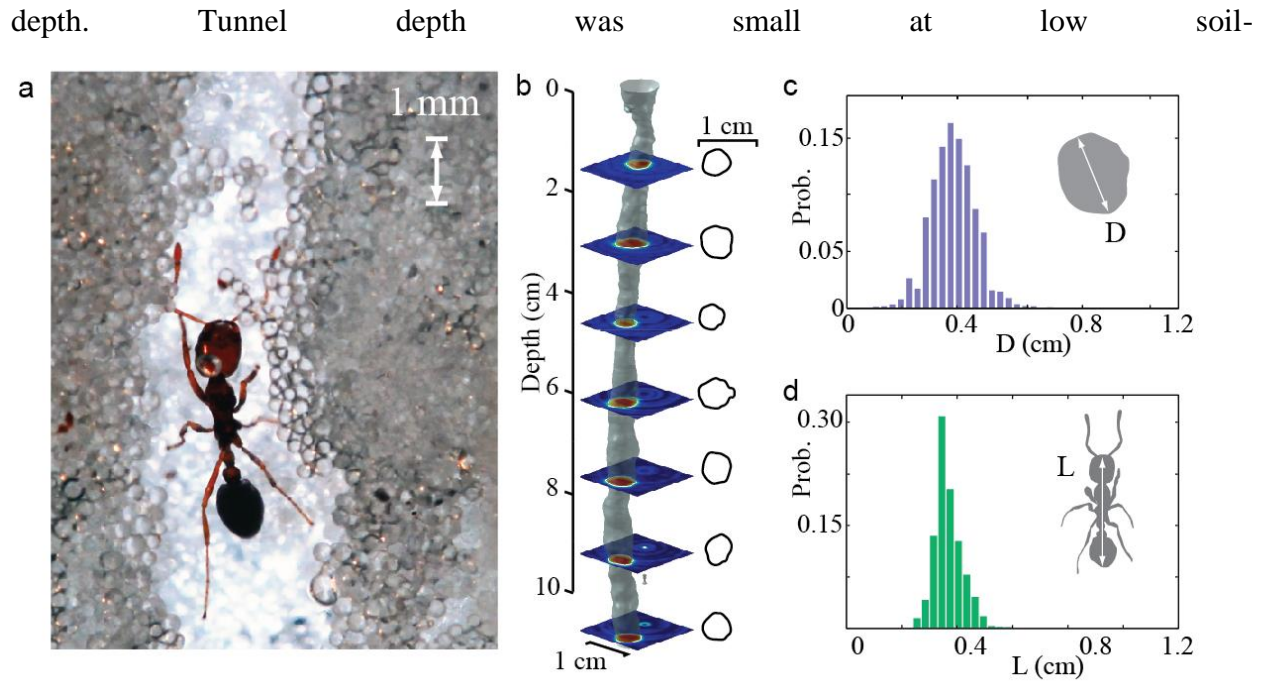

Fig. 1. Fire ants create and move through subterranean tunnels (Photo credit: Laura Danielle Wagner). a) Image shows fire ant worker climbing within an ant-constructed tunnel against a clear glass pane. b) X-ray CT scan reconstruction of a fire ant tunnel segment. c) Distribution of tunnel cross-sectional diameter, D. In A–C, the substrate consists of wetted 250 micron approximately spherical glass particles. d) Distribution of ant body length, L, (measured from tip of head to tail of gaster) in laboratory climbing experiments.

Importantly, however we found no significant effect of soil moisture ($F_{6,106} = 1.06$, p = 0.39), particle diameter ($F_{2,106} = 1.56$, p = 0.21), or the interaction of moisture and particle size ($F_{12,106} = 1.47$, p = 0.15), on the tunnel diameter (see Supplementary Material). Moreover, tunnels constructed in the laboratory and observed in X-ray CT were similar in diameter to tunnels found in natural fire ant nest mounds (4.4 mm;(47)), nest entranceways (3-4 mm,(46)), and incipient nests (3.1 ± 0.1 mm ;(48)), although tunnels deeper within natural nests may be larger in size (6.0 ± 3.0 mm; (47)). Our results demonstrate that during tunnel founding, fire ants show a relatively fixed behavioral program by building tunnels of approximately the same diameter in a variety of conditions. This suggests that the diameter of the tunnel could be important in fire ant locomotion.

*Tunnel size effects on the biomechanics of confined-climbing*

To investigate the biomechanics of locomotion within tunnels, we monitored fire ants climbing within ant-constructed tunnels within Quasi-2D arenas (Fig. 1a and SI Movies 1-3) and smooth cylindrical glass tubes (Fig. 2a-b). We tracked the position of ascending and descending ants freely trafficking between a foraging arena and nest through glass tubes of diameters, D = 1.0 - 9.0 mm (in increments of 1 mm). We will refer to these glass tubes as "glass tunnels".

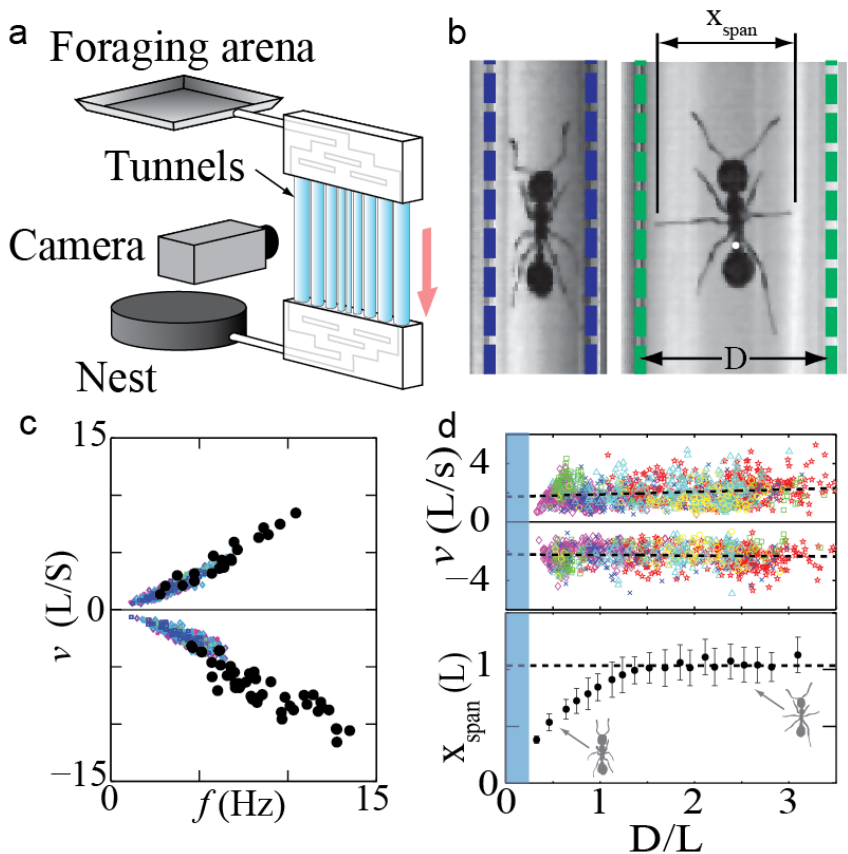

Fig. 2: Climbing posture and antennae use in glass tunnels. a) Schematic of climbing biomechanics experiment. b) Posture of ascending ant in a 6 mm diameter tunnel (left; in normalized units D = 1.36 L) and in a 2 mm diameter tunnel (right; D = 0.60 L). Left image shows posture variable, xspan, measured in experiment. Circle and white line indicate distance from touch-down location of limb to petiole. c) Stride frequency and speed relationship for glass tunnels (colored points) and ant-created tunnels (filled black circles). d) Top: Speed versus D/L. Color indicates colony. Dashed lines are linear fits described in the text. Blue box indicates the minimum predicted tunnel diameter an ant could fit in. Bottom: Lateral limb-span (mean ± s.d.) as a function of normalized tunnel diameter. Dashed line indicates constant limb-span of xspan = 1.04 ± 0.14 L independent of tunnel diameter.

We found that ants rapidly ascended (2.0 ± 0.8 L s$^{-1}$, N = 1621 ants) and descended (2.3 ± 0.7 L s$^{-1}$, N = 990 ants) in the glass tunnels (Fig. S4-5). The kinematic relationship between stride-frequency and speed (Fig. 2c) was fit by the function $v = ax^2 + bx$ for both the ascending ($a$ = 0.039 ± 0.003 L s; $b$ = 0.41 ± 0.01 L) and descending ($a$ = -0.018 ± 0.005 Ls; $b$ = -0.49 ± 0.02 L) climbs (Fig. S6-7). The speed-frequency relationship of ascent did not significantly differ among the ant-constructed tunnels and the glass tunnels of diameters, D = 0.3 – 0.4 mm (comparable to that of the self-constructed tunnels; F$_{2,361}$ = 1.8150, p = 0.1643). We did, however, find a small but significant difference in functional form of the speed-frequency relationship during descent ($F_{2,252} = 113.9$, $p < 0.001$). To test maximal performance within ant-constructed tunnels we induced an alarm response among the colony by exhaling into the tunnel entrance. Within ant-constructed tunnels ants rapidly descended ($6.9 \pm 2.1$ L s$^{-1}$; N = 21) and ascended ($4.1 \pm 1.8$ L s$^{-1}$; N = 45) at speeds greater than observed in the glass tunnels and surprisingly were able to move at speeds greater than 9 L s$^{-1}$ within the confined, simulated nest environment.

Tunnel diameter had a weak but significant effect on ascending speed (Top Fig. 2d), as a function of D/L ($v = m(D/L)+b$; F-test for non-zero slope, $F_{1,1619} = 63.132$, $p < 0.001$; $m = 0.17 \pm 0.04$ L$^2$ D$^{-1}$ s$^{-1}$, $b = 1.73 \pm 0.07$ L s$^{-1}$ ). During descent in tunnels, D/L did not have a significant effect on speed (F-test for non-zero slope, $F_{1,988} = 2.740$, $p = 0.10$). We thus hypothesized that the minimum tunnel diameter through which an ant can move is slightly larger than the animal's head width. Fire ant head width is $0.24 \pm 0.01$ L (49) and this sets the lower limit of the range of observable D/L values (Shaded blue box Fig. 2d). Both ascending and descending speeds near this lower limit (D/L < 0.5) sharply decreased (Top Fig. 2d) suggesting that only in the case of extreme confinement would we observe a strong effect of tunnel diameter on ascending or descending velocity. Overall, this suggests that ants move at a near constant upward and downward speed, over a wide range of tunnel sizes, while freely trafficking within the nest.

Tunnel diameter had a significant influence on climbing posture (Bottom Fig. 2d). Ants exhibited one of two stereotyped climbing postures: 1) within glass tunnels of D > L, ants adopted a sprawled posture in which mid-limbs were extended laterally away from the body (Fig. 2b, Right) and 2) within glass tunnels of D < L, mid-limbs were bent and pointed posteriorly (Fig. 2b, Left). We determined the critical tunnel diameter at which this postural transition occurred at by fitting the function $x_{span} = \begin{cases} k\left(\frac{D}{L}\right) & for\ D < D_c \\ c & for\ D > D_c \end{cases}$. We determined that in glass tunnels of diameter above $D_c = 1.03 \pm 0.01$ L the lateral limb-span, $x_{span}$, was independent of tunnel size ($R^2 < 0.001$) with mean value of $x_{span}$, determined from fit parameter c = $1.04 \pm 0.14$ L (Fig. 2d). In glass tunnels of diameter less than $D_c$, limb posture was altered by tunnel confinement and $x_{span}$ subsequently decreased (Fig. 2d). For comparison to ant-created tunnels, excavated tunnel diameter was D = $1.06 \pm 0.23$ L. Thus ants modify their limb position depending on tunnel size, but maintain approximately the same rate of ascent and descent. Furthermore ants climbing within tunnels they construct are capable of utilizing their spread-limb posture which may have implications for locomotor stability.

The alteration of the mid-limb posture in smaller tunnels suggests that a transition occurs in the direction of locomotor force production by the mid-limb. In the sprawled posture, mid-limb tarsi contact-forces pull towards the body and the tarsal hooks and adhesive pads are likely engaged. In contrast, when the limb is in the compact posture, the limb pushes down and away from the body to generate forward thrust. In the compact posture, to generate thrust force, we hypothesize that the rows of 50-350 μm long spiny hairs along the limb (Fig. S8) are utilized to engage asperities in the climbing substrate and allow the limb to push. Such multifunctional limb design has been previously shown to aid in rapid locomotion on horizontal substrates through the engagement of spiny limb hairs with rough surfaces (50).

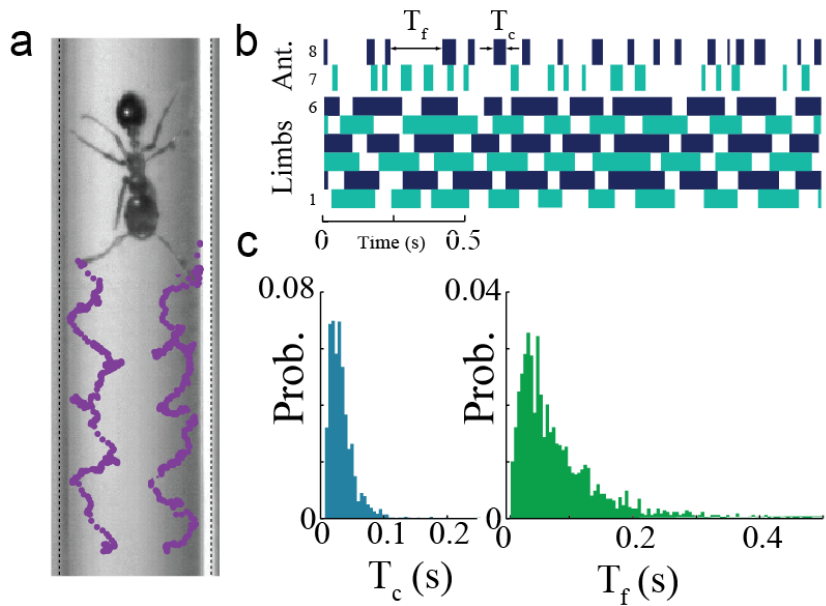

Fig. 3: Antennae use in confined locomotion. a) Image of ant descending in a tunnel with tracked position of antennae tips shown in purple. D=3 mm. b) Stepping and antennae contact diagram for a vertical descent in a tunnel. Light and dark blue highlight limbs that form alternating tripods during locomotion. Time of antennal contact, $T_c$, and time free, $T_f$, are highlighted. c) Probability distribution for both $T_c$ and $T_f$.

*Slip recovery through rapid jamming*

Fire ants possess a pair of elbowed antennae capable of a wide range of articulated motion about the head (Fig. 3a). While ascending and descending, ants rapidly placed antennae in contact with the tunnel walls (See SI Movie 1 and Fig. 3a). In the glass tunnels, antennae-wall contact time was $T_c = 29 \pm 23$ ms (Fig. 3c; N = 1840 contacts from 54 climbs) during head-first descent. The time between contacts was $T_f = 82 \pm 81$ ms (Fig. 3c). The rapid and repeated antennae-wall contact is important for tactile and chemo-sensing within the subterranean environment (35). However observations of ants slipping within glass and natural tunnels (SI Movie 4) led us to hypothesize that these sensory appendages could also have important biomechanical functions for climbing in confined spaces.

During high-speed ascent and descent in both glass and ant-constructed tunnels, ants exhibited slips that were rapidly corrected through antennae and limb contact with the tunnel surface (See Fig. 4 and SI Movies 1-3). Ants rapidly arrested short downward slips (in which the instantaneous downward velocity exceeded 15 mm s$^{-1}$) within $82 \pm 21$ ms (N = 456 slips among 54 individuals) within glass tunnels of all sizes. During head-first slips, antennae were placed against the tunnel wall prior to arrest in 92% of the observed slip-arrests (422 antennae contacts out of 456 slips). Excluding slips in which antennae began in contact with the wall, the time between slip onset and antennae-wall contact was $32 \pm 22$ ms (N = 265).

We briefly compare our observations of tunnel falling with the more extreme case of gliding among arboreal ants, in which ants in free fall can direct their motion during falls of hundreds to thousands of bodylengths (51). During aerial descent among canopy ants, gliding from a tree branch to a lower location on the tree aids in evasion from predators that may be on branches or on the forest floor. In the crowded and dark nest, long-distance directed aerial descent would be unsuccessful due to poor navigational ability and space constraints (lack of vision and tactile sensation from antennae). However, we hypothesize that the rapid slip-arrest we observe in high speed tunnel locomotion (See SI Movie 3 and Fig. 4) is an important mode of locomotion in confined environments, such that repeated slips or "micro-falls" can enhance rapid descent and maintain stability during climbing in tunnels..

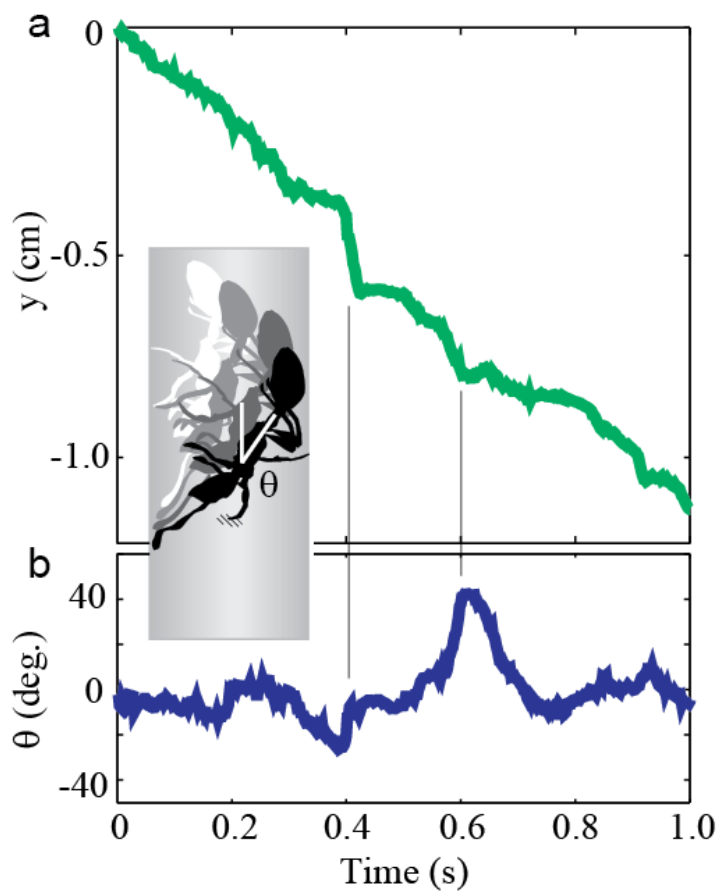

Fig. 4: Kinematics and perturbation-recovery during tunnel climbing. a) Vertical position of ant while descending (See Supplementary Movie 3). b) Body angle (θ) with respect to tunnel axis descending climb. Two slip-recovery events are highlighted by vertical gray lines. During slip events antennae and limbs are jammed against the wall and the body pitches into the tunnel face (illustration).

Our observations indicate that antennae are rapidly and readily used for slip-correction when climbing in confined space. In the case of larger slips, the antennae deformations also suggest that antennae provide significant mechanical support to the falling ant (SI Movie 4 and Fig. S9-11). Morphological adaptations to subterranean life are well documented(52); here we have observed for the first time that fire ant antennae—which are evolved from ancestral arthropod limbs (53)—retain partial functionality as locomotion appendages. Antennae can act effectively like 7[th] and 8[th] limbs to arrest falls.

Rapid fall arrest by bracing antennae against a tunnel wall relies on the ability to quickly jam limbs and body against opposing locations along the tunnel wall. Thus we hypothesized that the ability to rapidly arrest slips through body-jamming would be sensitive to tunnel diameter. To test this hypothesis we subjected ants climbing within glass tunnels to perturbations consisting of a rapid downward translation of the tunnels (Fig. 2a and SI Movie 5). Glass tunnels were mounted to a vertical air piston controlled through a computer. The piston translated the tunnels downwards 5.0 mm at which point the motion was stopped in less than 2.5 ms upon impact with the mounting plate. The final downward speed of the tunnels prior to impact was estimated to be 0.66 m/s; which suggest that ants were thus subject to a mechanical perturbation of ≈27g upon stopping. The perturbations employed in this experiment are substantially larger than what ants experience during jostling by neighbors in the natural environment. However, high speed perturbation-response experiments challenge the fastest neural response times of locomoting organisms, and thus help to determine the role of body kinematics and morphology in rapid locomotion stabilization(6, 10, 54, 55).

We found that 52% (1092 falls out of 2584 perturbations) of the perturbation experiments did not lead to ants being displaced from the tunnel wall (Fig. S12a). This indicates that the fire ant tarsi and adhesive footpads are robust to substantial perturbations, consistent with other measurements of the ant's adhesive strength (56-58). However, displacement from the tunnel surface did occur in 48% of experiments, and the outcome of perturbations was strongly influenced by the interaction of ant tunnel size.

We found that tunnel diameter, with respect to ant body length, had a significant effect on the probability to fall during a perturbation experiment, with smaller tunnels aiding in the ants perturbation resistance (Fig. S12B). The probability to fall during a perturbation increased from 36% to 73% as D/L increased from 0.4 to 3.4; the increase occurred over a narrow range around $D/L \approx 2.3$.. The high resistance to perturbation in tunnels of $D < 2.3$ L was likely due to the ability of fire ants to robustly engage surfaces. When climbing vertical planar surfaces, animals have to contend with gravity which, because the animal's center of mass is offset from the climbing surface, generates an overturning moment on the animal which must be overcome. In contrast, when climbing in small tunnels, ants may be able to minimize torque induced gravity on the body by placing limbs laterally against walls and thus keeping the center of mass in the same vertical plane as limb contact points.

Ants perturbed from the tunnel wall either arrested their fall within a vertical distance $\Delta y$, or fell to the tunnel bottom (Fig. 5b and SI Movie 5). Arrest distance, $\Delta y$, increased with increasing tunnel diameter normalized by bodylength, D/L (Fig. S13). The upper envelope of $\Delta y$ (dashed line in Fig. S13) increased linearly with a slope 67.0 ± 7.0 mm ($R^2 = 0.95$). This relationship can be understood through a kinematic argument: to arrest falls, ants extend limbs and antennae towards tunnel walls which are a further distance away within larger tunnels, and this results in longer fall distances in larger tunnels (See SI).

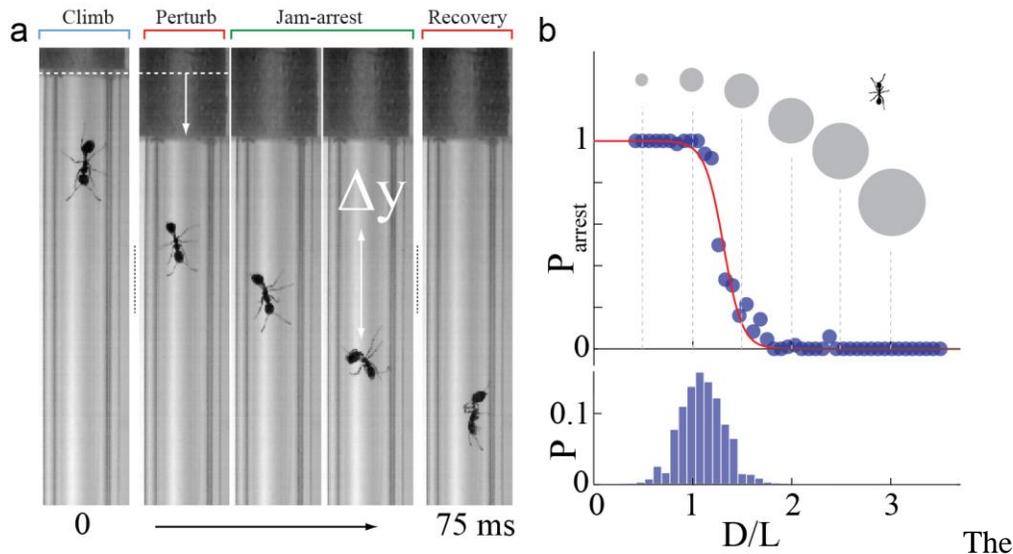

Fig. 5: Climbing perturbation experiment. a) Image sequence of perturbation and recovery (corresponding to perturbation 1 in Supplementary Movie 5). Left image is immediately prior to perturbation. Middle images show recovery which took place over 75 ms ($\Delta T$ = 30 ms for middle three frames). After perturbation recovery ant continued downward climb (Right image). b) (Top) Probability to arrest falls, $P_{arrest}$, versus D/L. Line is logistic fit described in text. Gray circles are tunnel diameter drawn to correct scale of ant illustration. (Bottom) Probability distribution of tunnel diameter in units of ant bodylength (L) for excavated tunnels (Data reproduced from Fig. 1d).

probability to arrest a fall, $p_{arrest}$, within a tunnel of size D/L decreased from 1 to 0 as D/L increased. We fit $p_{arrest}$ to a logistic function, $p_{arrest} = \frac{1}{1+\exp(\alpha(\frac{D-D_s}{L}))}$ (Fig. 5b) and found the cutoff tunnel diameter, $D_s$ = 1.31 ± 0.02 L ($\alpha$ = -10.54 ± 1.76), at which arrest probability decreased to below 50%. Within ant-constructed tunnels (of mean diameter 1.06 L; Bottom Fig. 5b) we predict that 93% of falls will be arrested. This demonstrates that ants display a high degree of climbing stability within tunnels of comparable size to those they create (1.06 L); however an increase in tunnel diameter by 50% reduced arrest probability to less than 5%.

We hypothesized that tunnel diameter would limit the ability to recover from falls through a, "jam-arrest" mechanism, within tunnels. Thus we expected that, $D_s$, was governed by morphological limitations of ant limb use. We measured the lateral limb-span, $x_{span}$, for free-falling ants and found that ants extended limbs maximally to a width of max($x_{span}$) = 1.33 ± 0.22 L independent of tunnel size when D > 1.3 L. This measurement is consistent with the typical mid-limb span of fire ants 1.31 ± 0.09 L reported in the literature (See SI and (49)) and suggests ants are extending limbs as much as possible to re-engage the tunnel wall while falling. In tunnels whose diameter exceeded the physical reach of the ants, D > 1.3 L, ants were unable to engage walls and the arrest probability decreased substantially (SI Movie 5).

We return to the digging experiments—in which groups of ants constructed tunnels—to understand how ant tunnel size relates to stability in confined spaces. The average diameter of tunnels created by ants across all excavation experiments was D = 1.06 ± 0.23 L. Thus it seems reasonable to assume that fire ants construct tunnels which facilitate rapid locomotion through the enablement of slip-recovery by antennae and limb jamming, without hindering limb kinematics. Many other factors are likely to influence equilibrium nest tunnel size. Traffic may be important in nest tunnel size determination as it is hypothesized that larger tunnels in the nest foraging network are due to higher traffic flow in these locations (47, 59). Further, food transportation requirements, ventilation, protection from flooding, and protection from invasion by predators or other brood raiding colonies can also influence nest structure (35). We hypothesize that the shape and size of tunnels at any time reflects the important environmental and biological factors influencing the colony at that time (38, 48). However, during incipient nest construction, such as after a flood, we expect that high speed locomotion and excavation are important to survival.

## Conclusions

We have shown that fire ants are capable of moving rapidly within their nest through the use of multi-functional limbs and antennae, which effectively engage surfaces within their constructed environment. We found that tunnel diameter had little effect on locomotion speed over a three-fold range of tunnel diameters, although body posture and limb use differed in different sized tunnels. We also discovered that fire ant antennae were effectively used as additional limbs during locomotion. Functionality of antennae as load-bearing, locomotor appendages was a surprising result, one which highlights the importance of studying locomotion within the context of the organism's natural environment. During locomotion, antennae were rapidly and repeatedly placed in contact with the tunnel surface for sensory feedback, however the antennae's multi-functional nature also means antennae may be the ant's first option in how to rapidly recover from missteps or slips. X-Ray CT indicated that fire ants constructed tunnels of appropriate size to enable utilization of the slip recovery mechanisms we observed in laboratory climbing experiments.

The ability for organisms to offload locomotion control to their environmental structures represents a new paradigm and a novel example of the integration of the organism's extended phenotype (the nest) for a locomotory purpose. We hypothesize that the construction of control surfaces suited to the locomotors body size and limb kinematics reduces locomotion control requirements within subterranean environments and may be a general feature of robust control within organism-engineered substrates such as tunnels, trails, or burrows. A universal scaling of burrow cross-sectional area with body-length(21)—sampled across a wide array of organisms varying by over six orders of magnitude by mass—provides evidence of the commonalities of locomotor constraints among subterranean animals. Thus the robust locomotor control strategies for subterranean environments we have described for fire ants may apply to a diversity of subterranean animals. We also expect that our biological discoveries will provide inspiration for, and simplify control in, collective robotic devices that will have to move within confined environments such as search and rescue zones. In addition, we propose that future robot teams could enhance survival in harsh terrestrial and extraterrestrial environments though collective construction of appropriately engineered shelters and nests.

## Methods summary
*Digging experiments*

We used a custom X-ray computed tomography (CT) system to observe tunnel excavation. Groups of 100-150 fire ant workers dug tunnels in 3.8 or 8.2 cm diameter chambers filled to a height of 12-15 cm with slightly polydisperse glass particles of diameter 50, 210, or 595 μm (Jaygo Inc., See SI Table 1 for particle size distribution). We varied water moisture content in the simulated-soil between 1-20% measured by mass. From CT-reconstructions we extracted the tunnel shape using the Chan-Vese active contours method(60). We measured the effective tunnel diameter, D, as the maximum of the distance transform of the tunnel cross section (See SI).

*Climbing experiments*

Climbs in ant-constructed tunnels were observed in quasi two-dimensional arenas, 27×34×0.3 cm in size, filled with wetted granular material as described in (61). Ants climbed between a nest and foraging arena through glass tunnels of diameter, D = 1 - 9 mm (in increments of 1.0 mm) and length 107.0 mm (Technical Glass, Ohio). Videos of climbing ants were recorded at a frame rate of 200 and 400 Hz (AOS imaging). To observe the falling response of ants within tunnels we performed a perturbation experiment in which a fixture holding the glass tunnels was mounted to a vertical, computer controlled air piston. The air piston accelerated the tunnels from rest 5.0 mm downwards over a time period of 0.15 s. Air piston activation was automated and triggered by ant movement which in turn triggered the capture of high speed video. All perturbed and unperturbed climbing experiments were performed while ants freely trafficked between the nest-site and the foraging arena.

*Statistics*

In all experiments ant body length was measured from the base of the mandibles on the head to the tip of the gaster. Ant body length was measured by selecting points in Matlab. Statistical tests were performed in Matlab and JMP (SAS Software). Analysis of variance was used for comparisons among treatments. In digging trials we treated colony and date as random factors in an analysis of variance (JMP). For comparing the statistical significance of nonlinear regression models to data we used the method described in Motulsky (62). All results are reported as mean ± standard deviation.

**Acknowledgements** Funding support was provided by National Science Foundation Physics of Living Systems Grant No. 095765 and the Burroughs Wellcome Fund.

**Author Contributions** N.G. designed the study and carried out experiments, analyzed data, and wrote the manuscript. D.M. carried out experiments and analyzed X-Ray CT data. M.A.D. designed the study and wrote the manuscript. D.I.G designed the study, directed the project and wrote the manuscript.

**Figure legends**

**Fig. 1**. Fire ants create and move through subterranean tunnels. a) Image shows fire ant worker climbing within an ant-constructed tunnel against a clear glass pane. b) X-ray CT scan reconstruction of a fire ant tunnel segment. c) Probability distribution of tunnel cross-sectional diameter, D. d) Probability distribution of ant body length, L, (measured from head to gaster) in laboratory climbing experiments.

**Fig. 2:** Climbing posture and antennae use in glass tunnels. a) Schematic of climbing biomechanics experiment. b) Posture of ascending ant in a 2 mm diameter tunnel (left; D = 0.60 L) and in a 6 mm diameter tunnel (right; in normalized units D = 1.36 L). Right image shows posture variable, $x_{span}$, measured in experiment. c) Stride frequency and speed relationship for glass tunnels (colored points) and ant-created tunnels (filled black circles). d) Top: Speed versus D/L. Color indicates colony. Dashed lines are linear fits described in the text. Blue box indicates the minimum predicted tunnel diameter an ant could fit in. Bottom: Lateral limb-span (mean ± s.d.) as a function of normalized tunnel diameter. Dashed line indicates constant limb-span of $x_{span} = 1.04 \pm 0.14$ L independent of tunnel diameter.

**Fig. 3:** Antennae use in confined locomotion. a) Image of ant descending in a tunnel with tracked position of antennae tips shown in purple. D=3 mm. b) Stepping and antennae contact diagram for a vertical descent in a tunnel. Light and dark blue highlight limbs that form alternating tripods during locomotion: (1) right-hind (2) right-mid (3) right-fore (4) left-fore (5) left-mid (6) left-hind. Time of antennal contact, $T_c$, and time free, $T_f$, are highlighted. c) Probability distribution for both $T_c$ and $T_f$.

**Fig. 4:** Kinematics and perturbation-recovery during tunnel climbing. a) Vertical position of ant while descending (See Supplementary Movie 3). b) Body angle (θ) with respect to tunnel axis descending climb. Two slip-recovery events are highlighted by vertical gray lines. During slip events antennae and limbs are jammed against the wall and the body pitches into the tunnel face (illustration).

**Fig. 5:** Climbing perturbation experiment. a) Image sequence of perturbation and recovery (corresponding to fall 2 in Supplementary Movie 5). Left image is immediately prior to perturbation. Middle images show recovery which took place within 75 ms (ΔT = 30 ms for middle three frames). Ant continued descending after perturbation recovery (Right image). b) (Top) Probability to arrest falls, $P_{arrest}$, versus D/L. Line is logistic fit described in text. Gray circles are tunnel diameter drawn to scale of ant illustration. (Bottom) Probability distribution of tunnel diameter in units of ant bodylength (L) for excavated tunnels (Data reproduced from Fig. 1d).

**Supplementary material**

**Supplementary movies**

**Movie S1:** Video of rapid descent in an ant constructed tunnel.

**Movie S2:** Video of rapid ascent in ant constructed tunnel. During the middle of the climb ant slips and falls backwards but its motion is rapidly arrested.

**Movie S3:** Rapid descent of an ant in an ant constructed tunnel subject to two slip-arrests. Tracked body position and orientation are shown with video and correspond to figure 3.

**Movie S4:** Nine slip-arrests observed in glass tunnels of diameter D = 0.3 − 0.4 cm. All falls are headfirst and antennae are utilized as $7^{th}$ and $8^{th}$ limbs to arrest fall.

**Movie S5:** Video illustrating perturbation experiment and stable and unstable falls.

**Supplementary methods**

**Ant collection and care**

The *S. invicta* colonies were collected during the spring of 2012 from roadsides outside of Atlanta, GA. Nests were excavated and transported to the laboratory and ants were separated from the soil using the water drip method (54). Colonies were housed in open plastic bins in a temperature controlled room with 12 hour on, 12 hour off lighting. Colonies were provided *ad libitum* water and insect larvae as food.

**Digging experiments**

The digging arenas were placed on a rotating stage controlled by a stepper motor (Lin Engineering) which was located 76 cm from a 110 kVp, 3 mA X-ray source. An image intensifier was located 103 cm from the source and a Phantom v210 camera (Vision Research) was used to visualize the X-ray images. Samples images were taken at angular increments of 0.9 degrees. We chose tunnels that were not adjacent to a wall (Fig. S1) and extracted the tunnel shape using the Chan-Vese active contours method (53). Tunnel properties were measured using the Matlab image morphology toolbox. We computed the distance transform of the tunnel shape using the Matlab command *bwdist* and considered the maximum value of the distance transform as the effective tunnel diameter.

*Digging experiment 1*

Groups of fire ant workers were challenged to dig tunnels in the laboratory. 8.2 cm diameter cylindrical containers were filled to a depth of 12 cm with a dry granular material of particle size $250 \pm 50$ μm (Jaygo Inc., Dragonite Soda Lime Glass beads, #5210). Arenas were first fully immersed in water to saturate the soil and then allowed to drain for 1 hour. Wet soil is known to induce digging in natural fire ant nests (33). Workers were introduced into the arena and were allowed to dig for 24 hours with a constant light source maintained above to stimulate digging. We evaluated tunnel the tunnel cross-section shape at various depths among 10 separate digging trials, each containing multiple tunnels, which resulted in 2,262 observations of tunnel diameter.

*Digging experiment 2*

In a second set of nest construction experiments we varied soil moisture content and particle size. We used collections of glass beads of diameter 50 μm, 210 μm, or 595 μm which were mixed with water and prepared at moisture contents of 1,3,5,10,15, and 20% (measured by mass). Supplementary Table 1 summarizes particle size distribution. Digging substrate was placed in a 3.8 cm diameter digging arena filled to a height of 14.5 cm. A 1 cm diameter plastic tube inserted into the center of the surface constrained the workers to initiate digging away from walls. We generated uniform compaction of the moistened media by sieving the wetted granular material through a mesh grid with 1 mm grid spacing using VTS 500 single vibrator system. Groups of 100 workers were introduced into the digging arenas and we evaluated tunnel shape in CT scans at 10, 15, and 20 hours. Eight separate colonies were tested at each particle size and moisture content combination resulting in 185 excavation experiments. We measured tunnel depth and cross-sectional shape at a depth of half the tunnel depth. We tested for the effect of particle size, water content, and the interaction (particle size)×(water content) using an analysis of variance in which colony and test date were treated as random effects.

## Climbing experiments

*Arena experiments*

Quasi two-dimensional arenas, 27×34×0.3 cm in size, were filled with the same wetted granular material as described in *Digging experiment 1* were prepared to allow for ant visualization during locomotion (See Gravish (54) for details). A group of 150 ants excavated in the simulated soil for 48 hours. We observed tunnel locomotion using a macro lens and a Phantom v210 camera, capturing video at 500 Hz. We encouraged high speed ascent and descent through ant-created tunnels by triggering an alarm response among the workers in which we exhaled gently into the nest entrance at the top surface.

*Glass tunnel climbing experiments*

We used a simulated-nest environment to study ant climbing in smooth glass tunnels in which we could view the interaction of all limbs and antennae with the climbing substrate (See Fig. 2). An enclosed, light-proof box which contained a wetted porous floor (plaster of paris) served as a nest, and housed 150-300 worker ants during the course of an experiment. The simulated-nest was connected to a foraging arena through a series of nine vertical observation tunnels ranging in inner diameter from $0.1 - 0.9$ cm in increments of 0.1 cm. Tunnels were 10.7 cm long and we observed a 9.6 cm length of them. Tunnels were illuminated by LED lights for visualization with a high-speed camera. *Ad libitum* water and food were provided in the foraging arena which encouraged worker traffic to and from the nest. A heat lamp was placed over the foraging arena to create a temperature gradient between the "above-surface" foraging arena and the "subterranean" simulated-nest. The simulated-nest and foraging arena setup encouraged ants to freely traffic within the tunnels and allowed us to observe tunnel climbing while performing a natural, unperturbed behavior.

Ant climbing posture was computed in Matlab in which we isolated the ant body from the stationary background using an active contours algorithm (53). We computed the vertically oriented bounding box of the ant-profile with the horizontal dimension of this box representing $x_{span}$. Climbing ants could be found at any angular location along the tunnel wall and thus we removed all runs in which ants were visualized from the lateral sides. Furthermore in measuring horizontal limb span we only included ant-postures in which the body axis measured from gaster to head deviated from the vertical by less than 10°. This resulted in 483,525 observations of climbing posture.

*Glass tunnel perturbation experiments*

To observe the falling response of ants within tunnels we performed a perturbation experiment. Glass tunnels were mounted to a vertical air piston maintained at 551 kPa and controlled through a computer. The piston's motion stopped upon impact with the mounting plate and vertical motion halted in less than 2.5 ms. We calculated that the final downward speed of the tunnels prior to impact was 0.66 m/s which suggests that ants were subject to a mechanical perturbation of 26.9 g upon stopping.

Activation of the air piston was controlled by a computer program which monitored motion in the upper portion of the tunnel region. When an ant was detected entering this region a relay was activated which controlled a high-speed solenoid that engaged the air piston. Simultaneously a trigger signal was sent to an AOS high speed camera which captured 2 second perturbation-response videos at 1024x1280, 400 fps and 500 μs exposure time. Analysis of perturbation experiments was performed using Matlab image analysis tools. Users determined fall-distance, ant-length, ant-orientation, fall-time, and fall code (successful arrest, no arrest, no fall) from the perturbation-response videos. We observed 2,268 perturbation-response experiments among worker ants from five of the six host colonies (B – F).

| 0-50 μm, Round. 80% | | 210-270 μm, Round. 80% | | 595-800 μm Round. 65% | |
|---|---|---|---|---|---|
| >60 μm | 0% | >420 μm | 0% | >1410 μm | 0% |
| >50 μm | 2% | >297 μm | 5% | >841 μm | 5% |
| >45 μm | 30% | >210 μm | 85% | >595 μm | 85% |
| >35 μm | 60% | >177 μm | 95% | >250 μm | 95% |
| >30 μm | 70% | | | | |
| <30 μm | 30% | | | | |

**Supplementary Table 1:** Distribution of particle size used in nest construction experiments. We used collections of glass beads of diameter 50 μm, 210 μm (Potters Industries Inc., Ballotini Impact beads, #6), or 595 μm (Potters Industries Inc., Ballotini Impact beads, #3). Rows indicate percentage of bulk mixture represented by that particle size or greater.

**Supplementary discussion**

**X-Ray computed tomography results data analysis**

We performed digging trials in 3D cylindrical containers of outer diameter 3.8 cm, and 8.2 cm. We identified tunnels that were not adjacent to the container wall and we extracted out their shape using an active contours algorithm as described in the Methods (Fig. S1). We characterized tunnel shape using two methods: 1) fitting an ellipse to the tunnel cross-section at different depths and 2) computing the distance transform of tunnel image mask, and multiplying the maximum by a factor of two. The distance transform of the image measures the nearest Euclidean distance to a tunnel wall at every pixel-location within the tunnel mask. Taking the maximum value of the distance transform for a given tunnel cross-section in effect estimates the "worst-case scenario" location for an ant to fall in that tunnel because that location is furthest away from tunnel surfaces. Since we are focused on locomotion stability we use the maximum of the distance transform as the metric for local-tunnel size and further refer to this as tunnel Diameter in the text.

Tunnels were primarily circular (Supplementary Figure 1) with major diameter $D_{maj} = 0.42 \pm 0.10$ cm and minor diameter $D_{maj} = 0.35 \pm 0.09$ cm, however there were significant cases in which tunnel shape deviated from a simple ellipse (Fig. S2) which warranted use of the distance transform technique. We note that since tunnels were primarily circular in cross-section (See ratio of major and minor axes in Fig. S2 e) the difference between tunnel size measured by ellipse fit or distance transform was small. Comparing the ellipse fit and image transform metric we find that the measured tunnel size in both cases have median values near unity (in units of ant body length, 1.04 L for image transform, and 1.15 L for ellipse). The distributions only differ substantially in the cases of larger tunnel diameters as expected from the case study in Supplementary Figure 2. Thus all references to diameter of ant constructed tunnels are determined using the image distance method.

We evaluated tunnel diameter and maximum depth from CT data after 10 and 20 hours of digging. We find that tunnel diameter did not increase over time (one sample t-test of relative change in diameter $\frac{D_{20h} - D_{10h}}{D_{10h}}$; t(34) = 0.7467, p = 0.4604) and instead incipient nests were enlarged through tunnel lengthening (one sample t-test of relative change in length $\frac{l_{20h} - l_{10h}}{l_{10h}}$; t(49) = 5.3644, p < 0.0001) consistent with a previous study of fire ant nest construction (54). We measured tunnel excavation in a diversity of idealized soil-substrates of varied moisture content and particle size (Fig.S3). We found that both particle size and soil-moisture had a significant effect on the maximum depth of tunnels over 20 hours. However, we found no significant effect soil-moisture content or particle size on tunnel diameter. Thus, the differences observed in tunnel depth indicate that soil-substrate properties did influence the digging ability of tunnel construction workers. However, the lack of significant change in tunnel cross-sectional morphology as a function of these varied simulated-soil conditions suggests that tunnel shape is actively being controlled for by the tunnel construction workers.

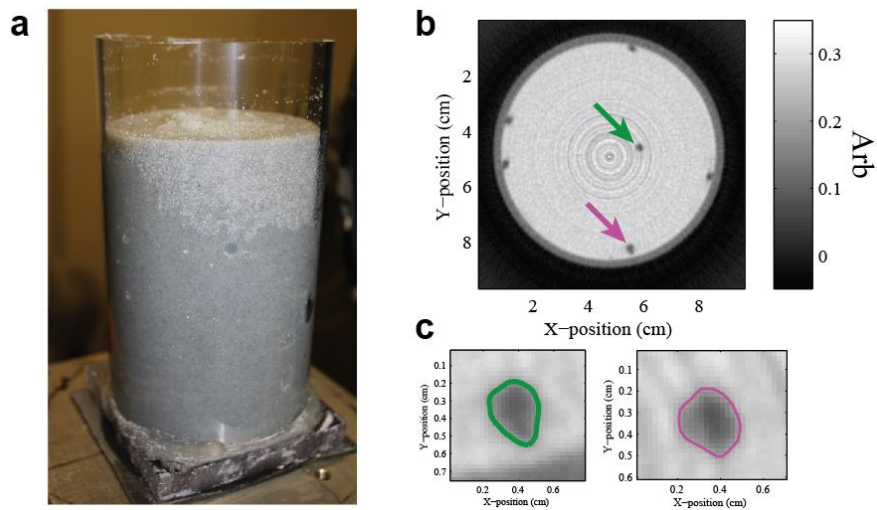

**Fig. S1:** Overview of the X-Ray CT digging trials. a) Digging arena consisted of a circular plastic, or aluminum container of diameters 8.2 or 3.8 cm, filled to a height of approximately 12-15 cm with a simulated soil of monodisperse 50, 210, or 810 μm diameter, wetted, glass beads. b) A horizontal cross-section from X-Ray CT reconstruction at a depth of 6.8 cm from the surface. Top and bottom arrow indicate two tunnels not adjacent to arena wall. Four other tunnels are present but against the tunnel wall. c) Cross-section of the top (left) and bottom (right) tunnels from (b) with extracted tunnel shape from active contours method shown as green (left) and purple (right) lines.

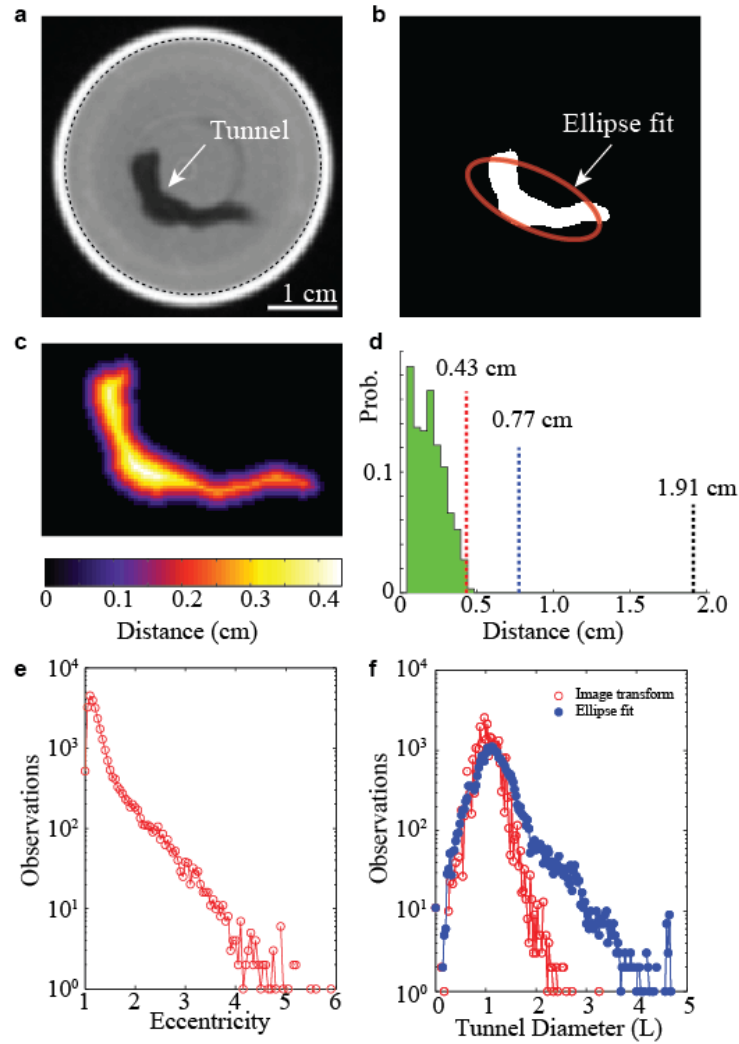

**Fig. S2:** Tunnel morphology analysis from X-Ray CT data. a) Tunneling experiment in a 3.8 cm outer diameter tube. Inner wall of digging arena shown as dashed line. An oblong shaped tunnel is highlighted in the center of the arena. b) Elliptical fit of tunnel shape. c) Euclidean distance image metric of tunnel shape. Color represents minimum distance of each pixel location to tunnel wall. d) Comparison of ellipse and image distance transform measures. Histogram of distance metric evaluated at all points in tunnel mask (green). Vertical dashed black line is major axis length from elliptical fit, dashed blue line is minor axis, and red line is maximum distance measured from distance transform. e) Distribution of eccentricity defined as major axis divided by minor axis from fitted ellipse fits. f) Distribution of tunnel diameter measured from distance transform and ellipse fit.

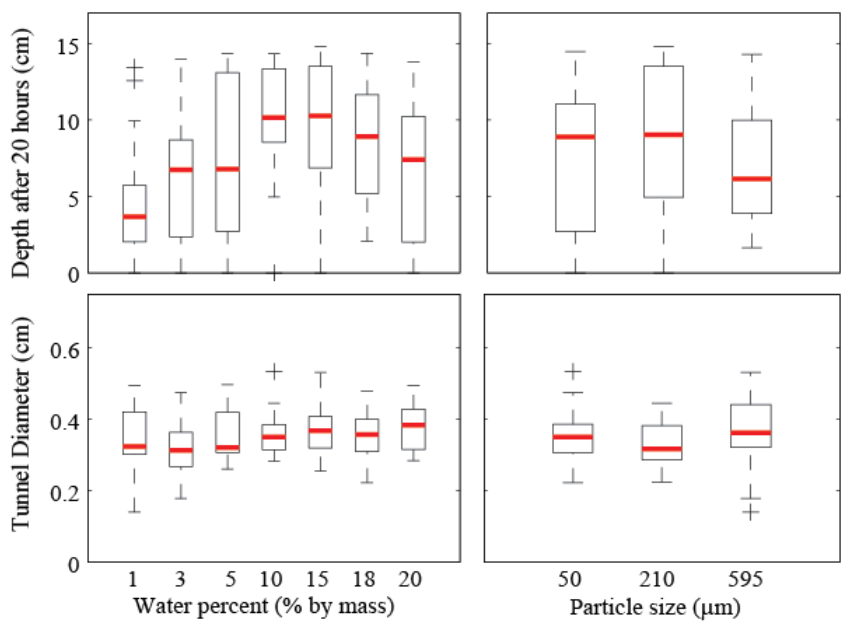

**Fig. S3:** Tunnel morphology as a function of particle size (left column) and water percent (right column). Box-plots of maximum tunnel depth after 20 hours are plotted along the top row and box-plots of tunnel diameter are plotted along the bottom row. We find that particle diameter and water percent statistically affected tunnel depth, but did not statistically affect tunnel diameter.

**Colony-level demographics from glass tunnel experiments**

We used ant groups drawn from six host colonies (A-F) for locomotion studies and five host colonies for perturbation experiments (B-F). We measured the body length of ants ascending and descending within tunnels and find that body length significantly differed among the host colonies (See Fig. S4a; $F_{5,2605} = 45.89$, $p < 0.0001$). We do not expect that the small differences (less than 15% difference between largest and smallest ant-length among colonies) among workers influenced any of the biomechanics results we present in this study. Furthermore all results are normalized by ant length to reduce possible variance due to differences in worker size. Worker size distribution within a fire ant colony varies as the colony ages with older colonies having larger workers (33). Thus the differences in worker size likely reflect the variance in host colony age.

We also observed significant differences in climbing speeds among the different colonies (Fig. S4b). Host colony had a significant effect on ascending climbing speed ($F_{5,1615} = 33.2$, $p < 0.0001$). The difference in speed between the fastest and slowest mean speeds among colonies was 28%. We also observed a significant effect of colony on descending climbing speeds (See Fig. S4c; $F_{5,984} = 6.06$, $p < 0.0001$) however the difference between the fastest and slowest mean speeds observed was only 12% in the case of descending. The difference in speed we observed may indicate different propensities to forage or explore among the colonies (observed tunnel locomotion consisted of workers moving between nest and foraging arena).

Lastly we observed small differences in the stability onset in $p_{arrest}$, measured in perturbation experiments. All colonies exhibited a transition from 100% arrest probability in small tunnels (D/L slightly larger than unity) and 0% arrest probability in large tunnels (D/L greater than 3). We characterize the stability onset as the parameter $\mu$ from the logistic fit function of arrest probability $p_{arrest} = \frac{1}{1+\exp(\alpha(\frac{D}{L}-\mu))}$. We find that $\mu$ varied with colony with fit value and 95% confidence intervals given in Supplementary Table 2 below. As can be seen in Fig. S5 all colonies exhibited a stability transition between 1.21 and 1.56 D/L evaluated over a range of D/L = 0.41 - 5.17.

| Colony | Fit (CI) |
|--------|----------|
| B | 1.209  (1.167, 1.251) |
| C | 1.56  (1.537, 1.583) |
| D | 1.294  (1.263, 1.324) |
| E | 1.485  (1.398, 1.572) |
| F | 1.337  (1.274, 1.4) |

**Supplementary Table 2:** Individual colony fit parameters for logistic fit of $p_{arrest}$.

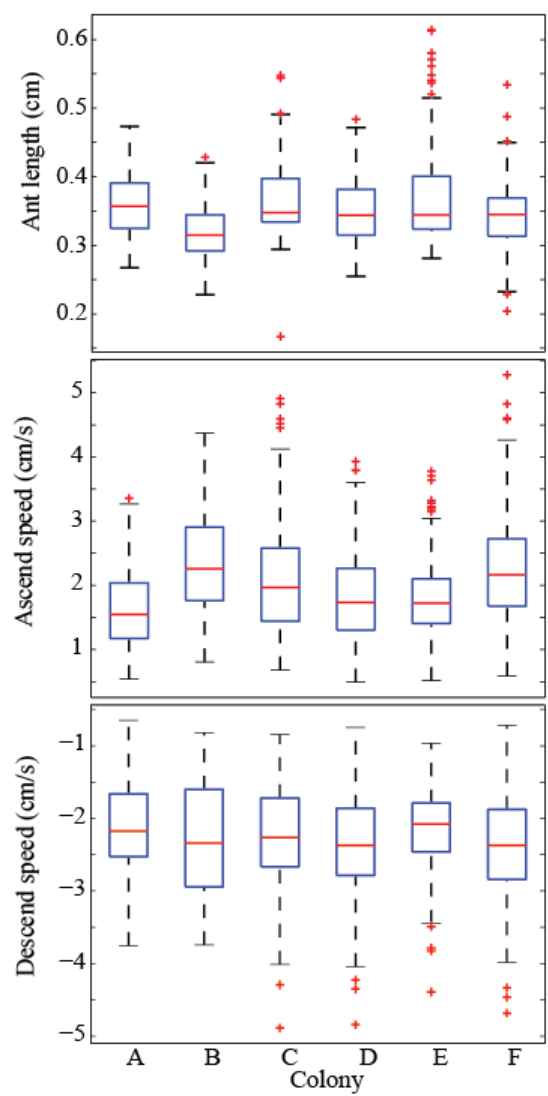

**Fig. S4:** Colony level demographics and speeds for the six colonies used in experiments. Red line is median value, blue box indicates 25% and 75% quartiles, and black dashed lines highlight maxima and minima. Red squares are outliers. a) Ant size distribution of different colonies. b) Descending speed of different colonies. c) Ascending speed of different colonies.

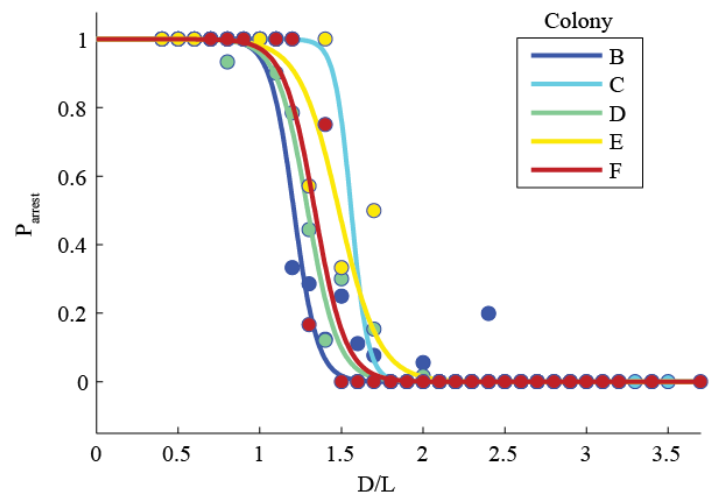

**Fig. S5:** $p_{arrest}$ with logistic fit for all colonies. Colonies denoted by symbol and line color as shown in legend.

**Comparison of locomotion kinematics in glass and ant-created tunnels**

The ascending and descending speed-frequency relationships in both ant-constructed and glass tunnels were similar (Fig. S6). In all cases speed increased with increasing stride frequency and this relationship was well described by a quadratic equation. We quantitatively compare the speed-frequency relationship between ant-constructed and glass tunnels to determine if ants modulate their gross climbing kinematics as a function of substrate. We fit functions of the form $v = ax^2 + bx$ to the speed-frequency kinematic relationship. To test for significant differences in climbing kinematics among ant-constructed tunnels and glass tunnels we use the method described in Reference (55) in which we compare the degrees of freedom and sums of squares of the individual fits and the data pooled together using an F-test. In the case of ascending climbs, comparison between glass (D = 0.3 - 0.4 mm) and natural tunnels we find no significant difference between the fit parameters from individually fit ant-constructed and glass tunnel data versus the pooled data ($F_{2,361} = 1.8150$, p = 0.1643). Ascending climbs were best fit with parameters a = 0.041 ± 0.003 cm s and b = 0.405 ± 0.015 cm ($R^2 = 0.95$). We find a statistically significant difference in speed-frequency relationship between glass (a = 0. 022 ± 0.01 cm s and b = 0.472 ± 0.045, $R^2 = 0.80$) and ant-constructed tunnels (a = -0.011 ± 0.013 cm s and b = 0.911 ± 0.140, $R^2 = 0.91$) in the case of descending. Individual fits of the glass tunnel and ant-constructed tunnels statistically describe the data better than when pooled ($F_{2,252} = 113.9$, p < 0.001). We note that the difference between ant-constructed tunnels and glass tunnels may be because ants move in a different locomotor mode (possibly through frequent slips) when descending at such high speeds as in ant-constructed tunnels.

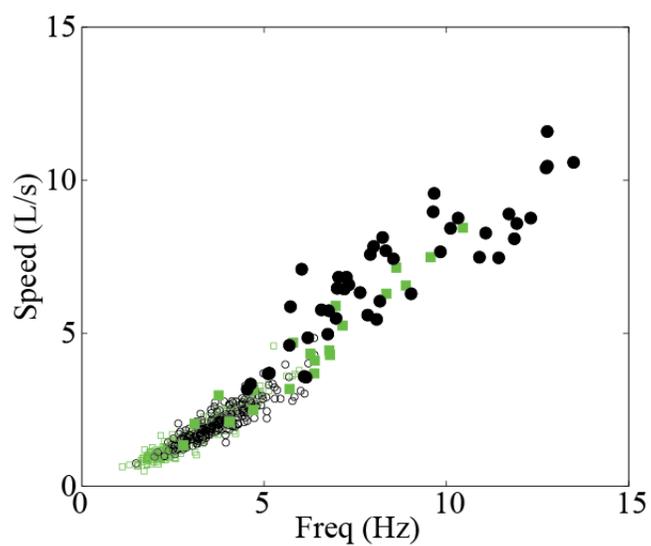

**Fig. S6:** Climbing speed-frequency realtionship in glass tunnels of diameter 3 and 4 mm (open symbols), and ant-constructed tunnels (closed symbols). Ascending climbs are shown as open and closed green squares. Descending climbs are open and closed circles.

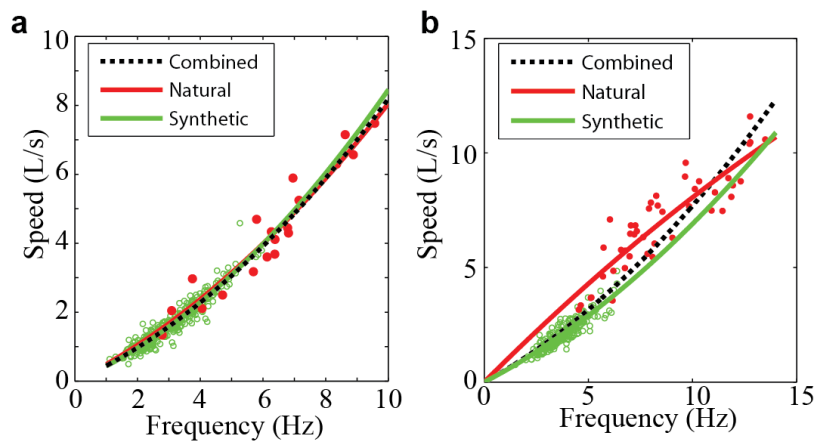

**Fig. S7:** Climbing speed-frequency relation in glass tunnels of diameter 3 and 4 mm (open symbols), and ant-constructed tunnels (closed symbols). a) Ascending speed-frequency relationship in ant-constructed and glass tunnels. Fit functions from natural, glass, and pooled data sets shown. b) Descending speed-frequency relationship in ant-constructed and glass tunnels. Fit functions from ant-constructed tunnels, glass tunnels, and pooled data sets (combined) shown.

**Mid-Limb morphology and posture statistics**

To quantify differences between the confined and unconfined locomotor postures are, we measured the fore-aft component of the anterior extreme position (AEP), the distance from petiole (the thin, central segment of the ant body) to limb touchdown, for the fore-limb, mid-limb, and rear-limb (Fig. S8). We isolated runs from two separate tunnel size treatments (small: $0.6 < D/L < 0.75$, N = 12 and Large: $2 < D/L < 2.25$, N = 15) that were at similar velocity ($2.5 \pm 0.18$ L/s) and measured limb touchdown locations. We observed a significant change in AEP for all three limbs (Fig. S8). Fore-limb touchdown distance significantly increased from $0.74 \pm 0.04$ L in the large treatment to $0.82 \pm 0.02$ L in the small treatment ($t_{25} = -5.7862$, $p < 0.001$). Mid-limb distance significantly decreased from $0.46 \pm 0.04$ L in the large treatment to $0.12 \pm 0.04$ L in the small treatment ($t_{25} = 22.55$, $p < 0.001$). Rear-limb distance decreased from $-0.53 \pm 0.04$ L in the large treatment to $-0.32 \pm 0.03$ L in the small treatment ($t_{25} = -13.0435$, $p < 0.001$). Thus all three limb pairs underwent a transition in kinematics between conditions of large tunnels which allow for full limb mobility, to small tunnels which constrain limb motion and require a postural alteration.

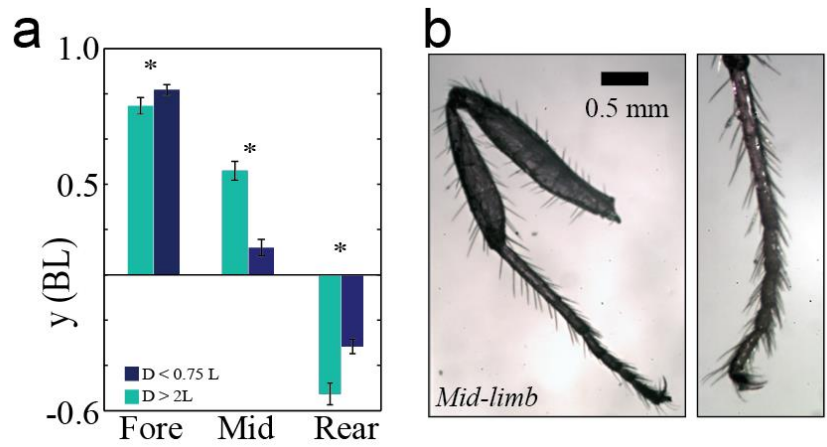

**Fig. S8:** Posture changes in two different sizes of tunnel. a) Fore-, mid-, and hind-limb touchdown distances in small (D < 1L) and large (D > 2L) tunnels. b) Visible light microscope images of *Solenopsis invicta* worker mid-limb showing the hairs that ling the limb pointing distally.

**Evidence of antennae use during falling**

We observed that in perturbed and unperturbed falls ants placed antennae against the tunnel wall and deflected or halted their center of mass motion about these antennae-tunnel contact points. Figure S9 shows an unperturbed fall in which the antennae are spread towards the wall at the initiation of a fall. In this example the ant fell a distance of approximately two bodylengths and arrested the fall through a combination of antennae and limb contact with the tunnel wall. Following this fall, the ant continued descending within the tunnel and the tracked points after the fall show the use of antennae while climbing down. In Fig. S10-11 we show eight additional examples of antennae use during unperturbed head-first falls. Video of each fall is shown in SI Movie S4.

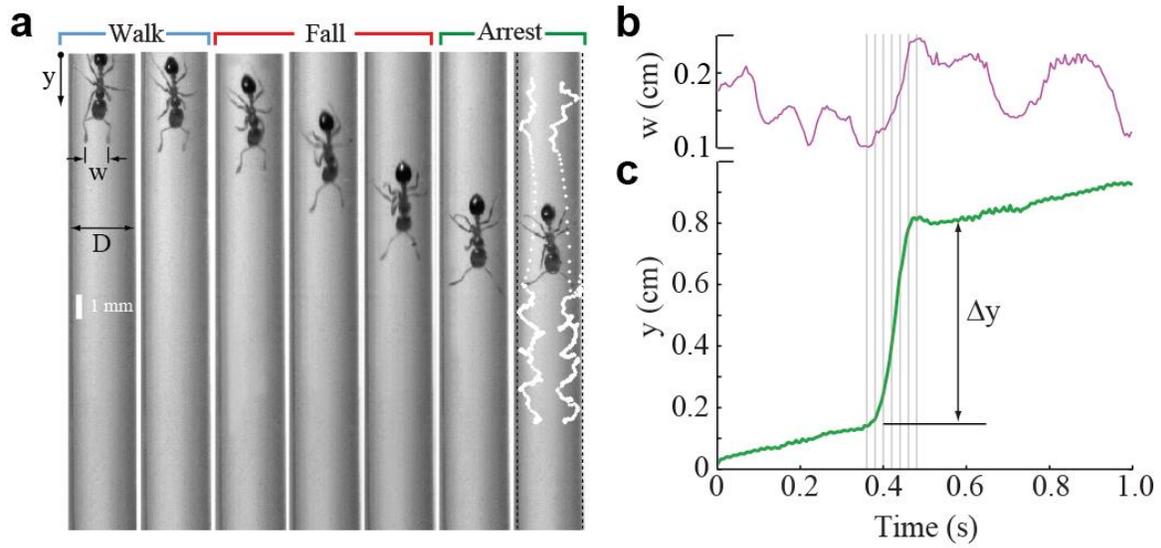

**Fig. S9:** Fall and arrest using antennae. a) Time-lapse images of an ant descending in a tunnel. Images are separated by 10 ms. White dots in last frame show tracked position of antennae prior to, during, and after fall. b-c) Instantaneous antennae width (top) and vertical position of ant body (bottom) during the fall-arrest. Grey lines indicate images in b. Δy indicates fall distance.

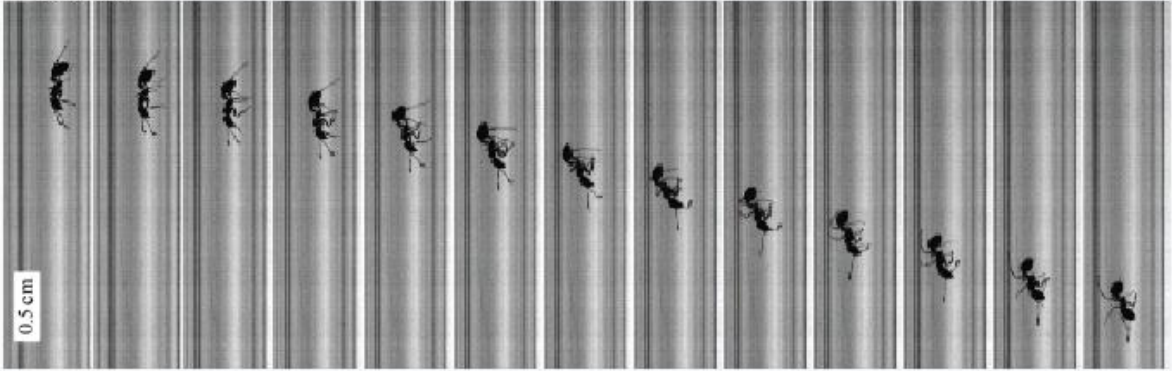

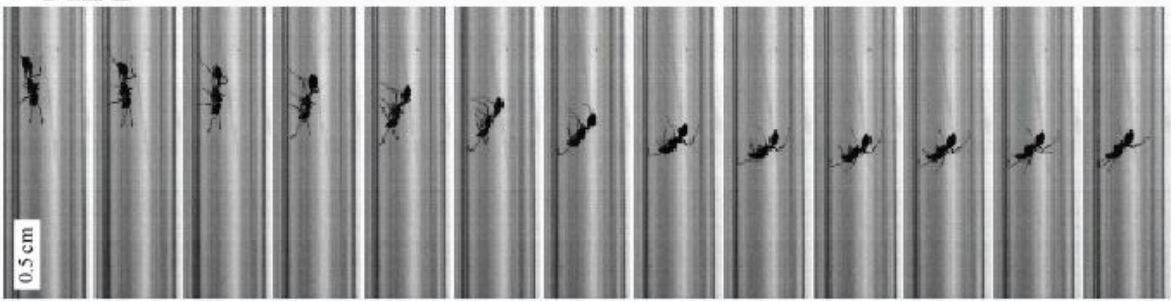

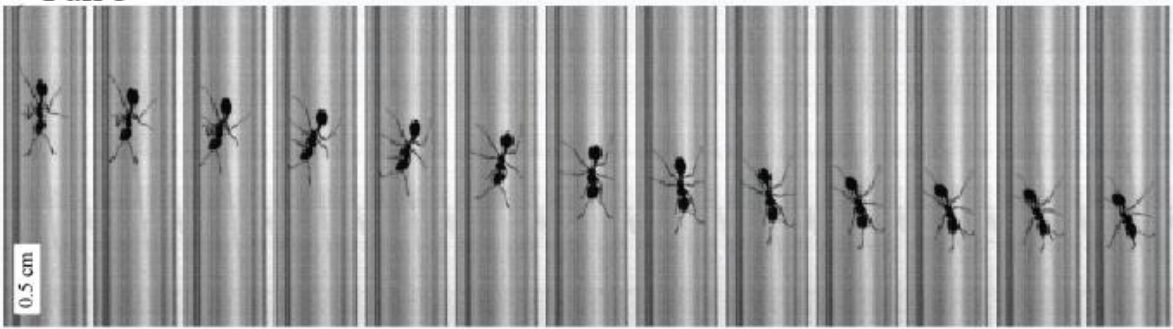

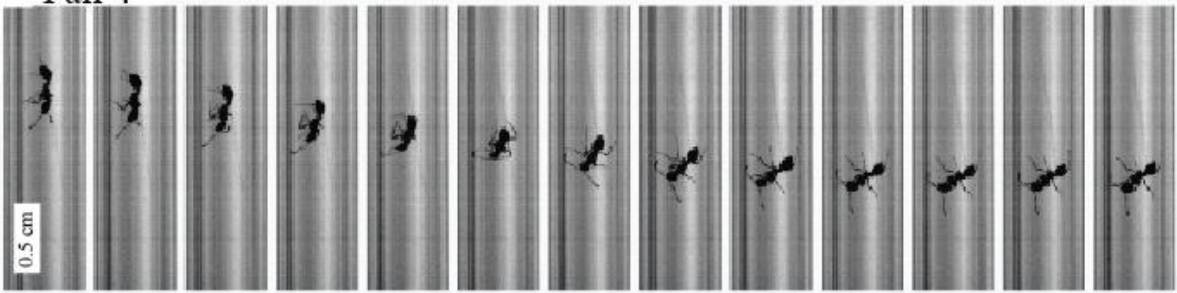

**Fig. S10:** Four unperturbed, head-first, fall-arrest sequences in glass tunnels. All falls illustrate use of antennae during fall-arrest. Images are separated by 5 ms time intervals. All falls in tunnel size D = 0.4 cm.

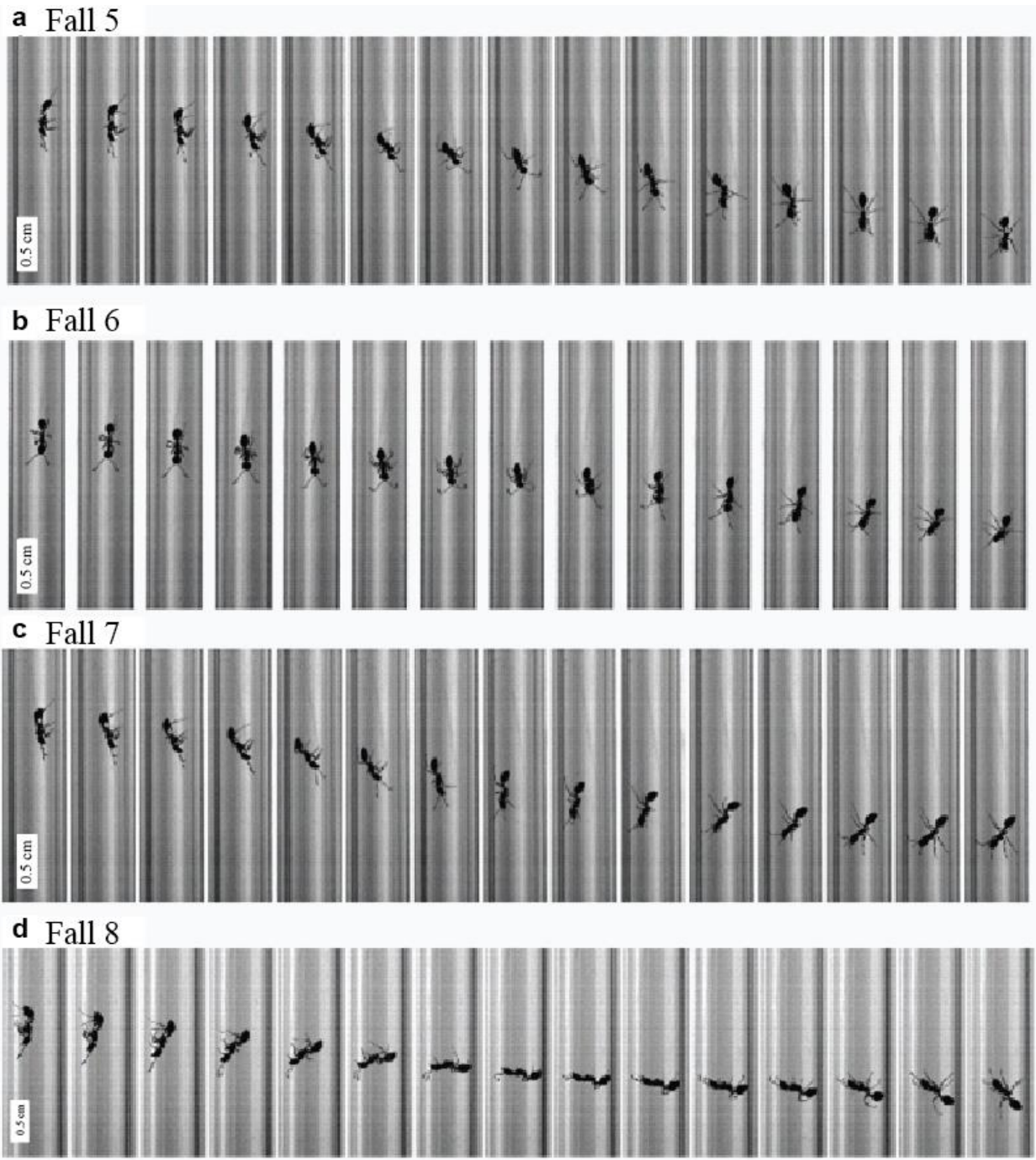

**Fig. S11:** Four unperturbed, head-first, fall-arrest sequences in glass tunnels. All falls illustrate use of antennae during fall-arrest. Images are separated by 5 ms time intervals. Tunnel size D = 0.4 cm (a), 0.4 cm (b), 0.4 cm (c), and 0.6 (d).

**Perturbation experiment statistics**

We found that 52% (1092 falls out of 2584 perturbations) of the perturbation experiments did not lead to ants being displaced from the tunnel wall (Fig. S12a). Tunnel diameter, with respect to ant bodylength, had a significant effect on the probability to fall during a perturbation experiment with smaller tunnels aiding in the ants perturbation resistance (Fig. S12b). We fit the fall probability with a logistic function $p_{fall} = a + \frac{b}{1+\exp(\alpha(\frac{D_f - D}{L}))}$ (a = 0.37 ± 0.05, b = 0.36 ± 0.09, α = 12.69 ± 17.95, $D_f$ = 2.28 ± 0.15). The probability to fall during a perturbation doubled from 36% when D < 2.28 L, to 73% when D > 2.28 L.

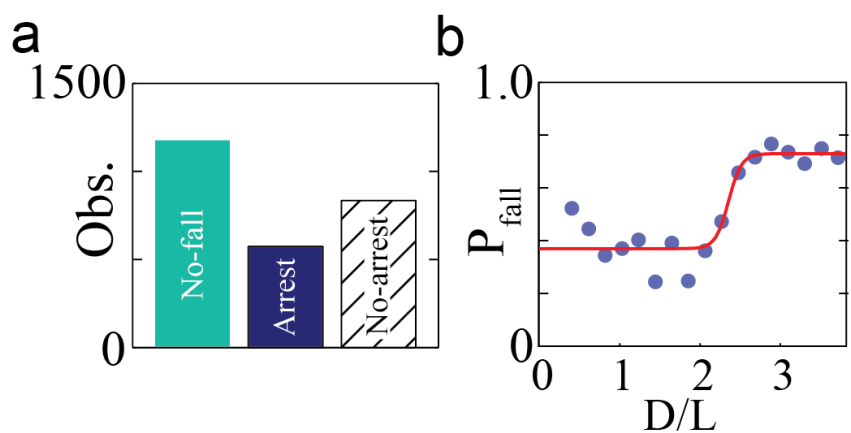

**Fig. S12:** Falling statistics. a) Number of observations from perturbation experiments. b) Probability to fall during a perturbation experiment as a function of D/L. Red line is logistic fit described in text.

## Perturbation experiment statistics

We measured the distance perturbed ants fell, $\Delta y$, in within tunnels of different diameter (Fig. S13). We found that fall distance increased with increasing tunnel diameter. To gain insight into this relationship we construct a simple model of fall-arrest in tunnels. We assume that falling ants must extend their limbs out to regain contact with the wall. We define The lateral limb distance from the body axis as $l$, and the distance the limb is from the wall as $\Delta l = D - l$ (we assume the ant is in the center of the tunnel). An ant falling from rest under gravity will accelerate downwards with a vertical position as a function of time given by $\Delta y = \frac{1}{2} g t^2$. We assume that the limbs are accelerated outwards from the body at a constant rate, $a$, which results in horizontal limb position $l = \frac{1}{2} a t^2$. Eliminating $t^2$ from the two equations we obtain the relation $\Delta y = \frac{g}{a} l$ which can be put in terms of tunnel diameter as $\Delta y = \frac{g}{a} (D - \Delta l)$. Thus we see that fall distance scales linearly with tunnel diameter and the upper envelope of arrest points is well approximated by a line.

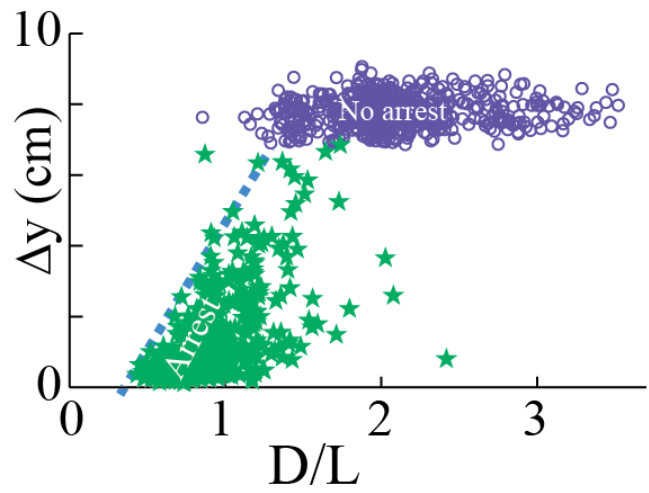

**Fig. S13:** Correlation between tunnel morphology D/L and fall distance. Fall distance, Δy, as a function of D/L. Purple circles indicate no arrest while green stars indicate arrest.

**Falling posture and lateral limb-span**

For each perturbation experiment we measured the lateral distance, $x_{span}$, that falling ants extended limbs and antennae to in the horizontal plane as a function of time (Fig. S14). Ants perturbed from tunnel walls extend their limbs and antennae laterally away from their body to re-engage contact with the tunnel surface during a fall. To determine the maximum lateral limb and antennae span ants display when falling we measured the maximum of $x_{span}$. We find that $max(x_{span})$ was limited by tunnel diameter in tunnels $D < 1.3$ L and was fit by a linear equation, $max(x_{span}) = aL$ ($a = 1.18 \pm 0.13$, $R^2 = 0.71$). In tunnels of $D > 1.3$ L $max(x_{span})$ was independent of tunnel size ($R^2 = 0.05$) with value $max(x_{span}) = 1.33 \pm 0.22$ L. These measurements indicate the limits of limb-antennae extension ants are capable of when attempting to arrest falls. The value of maximum limb extension equal to 1.33 L suggests that in tunnels above 1.33 L in diameter, we should not observe jamming mode arrest and successful arrest of falls should decrease. This is supported by the logistic fit parameters of $p_{arrest}$ in which the transition from successful arrest to unsuccessful arrest occurs at a critical tunnel size of $1.31 \pm 0.2$ L.

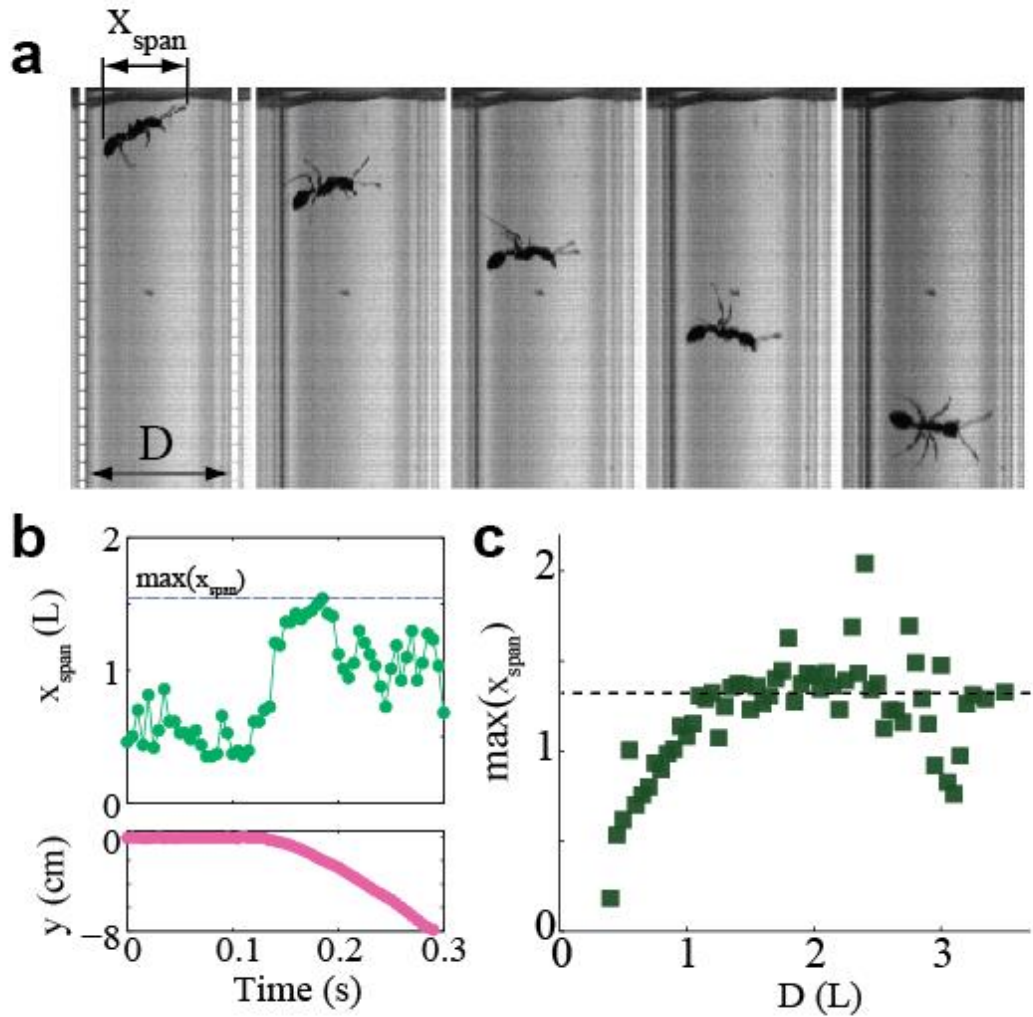

**Fig. S14:** Falling posture and arrest probability. a) Five images from a perturbed fall. Images are separated by 10 ms. During perturbed falls we measured the total horizontal span of limbs, antennae, and body, called $x_{span}$. b) $x_{span}$ and vertical position, y, plotted versus time. Perturbation occurs at 0.1 s and vertical descent is accompanied by an increase in $x_{span}$. c) Maximum lateral limb-span [max($x_{span}$)] exhibited during fall as function of D(L). Limb-span is constant above D > 1.3 L with max($x_{span}$) = 1.33 ± 0.22 L (dashed line shown).

**Morphological measurements of fire ants to estimate stability-limits for tunnel arrest**

     Using data from Reference 45 we estimated the maximum limb and antennae span fire ants were able to reach while falling (Fig. S15). We estimate the antennae span ($d_1 = 0.86 \pm 0.08$ L) as two times the total antennal length (scape length + club length). We estimate mid-limb span, denoted simply as limb span ($d_2 = 1.31 \pm 0.09$ L), as two times the total leg length (femur + tibia + tarsi) of the mid-limb. Lastly we estimate the full length along the head-gaster body axis ($d_3 = 1.85 \pm 0.09$ L) as the antennae length + head length + alitonum length + hind-limb length.

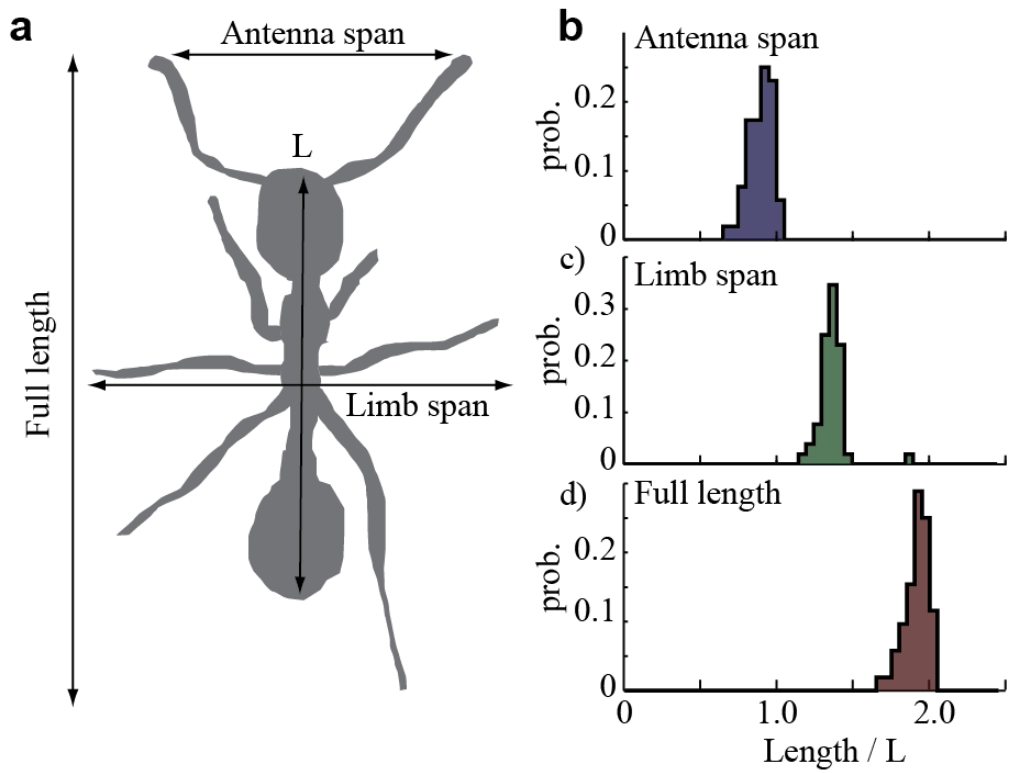

**Fig. S15:** Fire ant morphology data. Here we re-analyze the data reported from Tschinkel et al. (45) of morphogical measurements of the appendages and body segments of the fire ant *Solenopsis invicta*. a) We measure the antennae span, the mid-limb span, and the full length distance defined as antenna length + head length + alinotum length + rear leg length. b-d) Histograms of morphological measurements hypothesized to be relevant to tunnel arrest in units of body length.